\documentclass[aps,prb,twocolumn,showpacs]{revtex4}
\usepackage{tabularx,graphicx}\begin{document}
\title{Effect of assisted hopping on the formation of local moments
in magnetic impurities and quantum dots.}
\author{F. Guinea}
\affiliation{
Instituto de Ciencia de Materiales de Madrid,
CSIC, Cantoblanco, E-28049 Madrid, Spain.}
\date{\today}
\begin{abstract}
Assisted hopping effects in magnetic impurities and quantum dots
are analyzed. The magnitude of the assisted hopping term in 
a quantum dot in the limit of large level spacing is comparable
to other corrections induced by the 
electron-electron interactions. Assisted hopping leads
to differences between conductance peaks associated to the
same level, and, when the effect is sufficiently strong, to 
local pairing correlations.
\end{abstract}
\maketitle
\section{Introduction}
Electron-electron interactions lead to the formation of local
moments and to the Kondo effect in magnetic 
impurities in metals\cite{K64,A61},
and in quantum dots attached to metallic 
leads\cite{GR88,Getal98,Cetal98}. This effect is the
direct consequence of the Coulomb repulsion between
electrons which occupy the same quantum level of the
impurity or quantum dot.

If we assume that the main physical features of the
impurity (or the dot) are governed by a single quantum level,
the leading correction to the intralevel Coulomb interaction
is an assisted hopping term. This interaction induces a dependence of
the coupling of the impurity (or the dot) and its
environment on the occupancy of the level\cite{H93}.

In the following, we estimate the magnitude of
such a term for models of large atoms and quantum dots, 
and analyze its effects 
on the formation of a local moment on the impurity, or quantum dot.
For bulk systems, the inclusion of an assisted hopping term in the
electronic hamiltonian favors the existence of pairing 
correlations\cite{HM91}. In the case of an impurity, this tendency
towards local pairing quenches the local moment, and, for
quantum dots, it can lead to asymmetries in the conductance
of peaks associated to the same level, and even to
an enhancement of the conductances at low temperatures.

The model is described in the following section IIA. Simple estimates
of the magnitude of the assisted hopping term for quantum dots
are discussed in section IIB. A mean field analysis is given in
sections IIIA and IIIB. Extensions of the mean field approach are 
presented in section IIIC. Section IV contains the main conclusions.
\section{The model}
\subsection{The hamiltonian.}
We analyze a single quantum level, associated to the creation
operator $d^\dag_s$, where $s \equiv \uparrow , \downarrow$ is the
spin. This level is coupled to a continuum of non interacting
electrons, described by operators $\sum_k c^\dag_{k s}$. For simplicity, we assume that
a single channel in the environment interacts with the impurity\cite{channel}.  
The hamiltonian is:
\begin{widetext}
\begin{equation}
{\cal H} =
\sum_{k,s} \epsilon_{k} c_{k,s}^\dag c_{k,s}
+ \epsilon^0_d n_d + U n_{d \uparrow} n_{d \downarrow}
+ ( V - \Delta V n_{d \uparrow} ) \frac{1}{\sqrt{\cal V}}
\sum_k c^\dag_{k \downarrow}
d_\downarrow + ( V - \Delta V n_{d \downarrow} )
\frac{1}{\sqrt{\cal V}} \sum_k c^\dag_{k \uparrow} d_\uparrow + h. c.
\label{hamil}
\end{equation}
\end{widetext}
where ${\cal V}$ is the volume of the system, and:
\begin{eqnarray}
n_{d \uparrow} &= &d^\dag_\uparrow d_\uparrow \nonumber \\
n_{d \downarrow} &= &d^\dag_\downarrow d_\downarrow \nonumber \\
n_d &= &n_{d \uparrow} + n_{d \downarrow}
\end{eqnarray}
The density of states of the conduction band at the Fermi level
is $N ( \epsilon_F )$. We assume that
$V N ( \epsilon_F ) , \Delta V N ( \epsilon_F ) \ll 1$.
Without loss of generality, we will set $\epsilon_F = 0$,
and define $N_0 = N ( \epsilon_F )$.

When $\Delta V = 0$ we recover the standard model proposed by Anderson
for the study of the formation of local moments in metals\cite{A61}.
This model is determined by three parameters, the Coulomb repulsion
$U$, the position of the level $\epsilon^0_d$, and its width,
$\Gamma = V^2 N_0$. The model has electron-hole symmetry around
 $\epsilon_d = \epsilon^0_d - U / 2$.

The assisted hopping term in eq.(\ref{hamil}) is  the next leading 
interaction in terms of $d$ operators which can be added to
the Anderson hamiltonian, assuming the same truncated 
Hilbert space. As discussed later, it can be derived from
intradot interactions only.
The hamiltonian (\ref{hamil}) intorduces an additional
dimensionless parameter, $\Delta V / V$. For small
atoms, $\Delta V / V$ is a number of order unity\cite{H93}. 
The inclusion of this term in the hamiltonian breaks the
electron-hole symmetry of the initial Anderson hamiltonian.
\section{Calculation of $\Delta V$.}
\subsection{Definition of $\Delta V$.}
We assume that the assisted hopping term in eq.(\ref{hamil})
arises from interactions which involve electronic levels
within the dot only.  When the hamiltonian is truncated to
the Hilbert space defined by a single state within the dot,
and the leads, the electron-electron interactions within
the dot can induce, among others, a term of this type.

Our initial hamiltonian is:
\begin{eqnarray}
\cal{H} &= &{\cal H}_{dot} + {\cal H}_{lead} + {\cal H}_{tunn}
\nonumber \\
{\cal H}_{dot} &= &\sum_i \epsilon_i c^\dag_i c_i +
\sum_{ijkl} h_{ijkl} c^\dag_i c_j c^\dag_k c_l \nonumber \\
{\cal H}_{lead} &= &t \sum_{n=0}^\infty \bar{c}^\dag_n \bar{c}_{n+1}
+ h. c. \nonumber \\ 
{\cal H}_{tunn} &= &\sum_i V_i c^\dag_i \bar{c}_0 + h. c.
\label{dot}
\end{eqnarray}
We have assumed that the lead contains a single
channel. The index $i$ labels all quantum numbers of the states
within the dot, including spin. 
We are interested in transitions where the charge
state of the dot changes from $N$ to $N+1$, and from $N+1$ to $N+2$.
In the absence of interactions, they correspond to the filling
of a particular level, which we denote $i$. When the charge
state of the dot is $N$, all levels $j$ such that $j<i$ are occuppied.
We use as basis for the electronic states in the dot the Hartree-Fock
wavefunctions defined when for charge state $N$. Rearrangements of
the electronic levels imply that the corresponding wavefunctions
for charge states $N+1$ and $N+2$ are different\cite{WMG99}.
Typically, the tunneling between the dot and the lead takes
place in a region of atomic size at some point of the surface
of the dot, ${\bf \vec{r}}_0 $. Then, $V_i \propto
\Psi_i ( {\bf \vec{r}}_0 )$, where $\Psi_i ( {\bf \vec{r}} )$ is the
wavefunction of state $i$.

We now assume that the broadening of the levels due to the
coupling to the leads, $\Gamma = \langle V_i^2 N_0 \rangle$ is smaller
than the mean level spacing, $\Delta$. Then, the effective tunneling
can be estimated from the ground state wavefunctions of the dot for
different charge states:
\begin{equation}
V_{eff}^{N \rightarrow N+1} = \langle N+1 | \sum_j V_j c^\dag_j
| N \rangle
\label{overlap}
\end{equation} 
and the difference between adding two electrons to state $i$,
and adding  one is:
\begin{equation}
\Delta V = V_{eff}^{N+1 \rightarrow N+2} -
V_{eff}^{N \rightarrow N+1}
\label{deltav}
\end{equation}
\subsection{Perturbative analysis.}
For large dots, the interaction terms decrease with the size
and conductance of the dot\cite{BMM97,A00,ABG02}, so that
a perturbative calculation of $\Delta V$ is possible.

As mentioned above, we use the electronic basis which diagonalize
the Hartree-Fock approximation to
${\cal H}_{dot}$ in charge state $N$, which we denote ${\cal H}_{HF}^N$.
We can define ${\cal H}_{int} = {\cal H}_{dot}
- {\cal H}_{HF}^N$, and the wavefunctions:
\begin{eqnarray}
| N \rangle_0 &= &| N \rangle_{HF} \nonumber \\ 
| N + 1 \rangle_0 &= &c^\dag_{i \uparrow} | N \rangle_{HF} \nonumber \\
| N + 2 \rangle_0 &= &c^\dag_{i \uparrow}
c^\dag_{i \downarrow} | N \rangle_{HF}
\label{wavefunctions}
\end{eqnarray}
where $| N \rangle_{HF}$ is the ground state of
${\cal H}_{HF}^N$ (note that $| N + 1 \rangle_0$ and
$| N + 2 \rangle_0$ are not the Hartree-Fock approximations to
the wavefunction of the dot with charge $N+1$ and $N+2$).

\begin{figure}
\resizebox{6cm}{!}{\includegraphics[width=7cm]{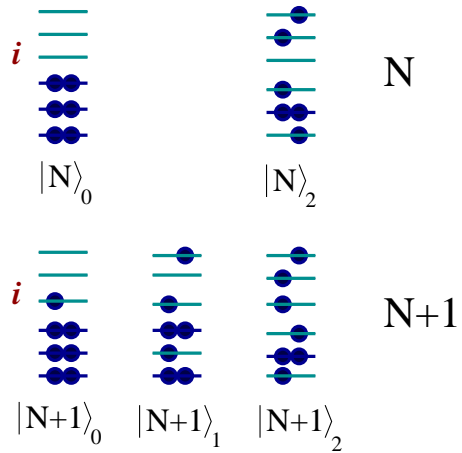}}
\caption{Sketch of the zeroth order wavefunctions, and
corrections with one and two electron-hole pairs for charge
states $N$ and $N+1$. See text for details.}
\label{wave}
\end{figure}

The interaction hamiltonian, ${\cal H}_{int}$, when
acting on $| N \rangle_0$ induces two electron-hole pairs.
The occupation of level $i$ in
wavefunctions $| N + 1 \rangle_0$ and
$| N + 2 \rangle_0$, and the induced
interaction between this electron and the rest
implies that the basis which
diagonalizes ${\cal H}_{HF}^N$ is not longer optimal. 
Then, ${\cal H}_{int}$ acting on these wavefunctions
leads to excited states with one and two electron-hole pairs.
A sketch of the corrections to the wavefunctions is shown
in Fig.[\ref{wave}].

Matrix elements which involve
one and two electron-hole pairs have a different dependence
on the size of the dot (see below). We will calculate
$\Delta V$ to second order in the matrix elements associated
to one electron-hole pair, ${\cal H}_{int,1}$,
and to first order in the matrix elements
involving two pairs, ${\cal H}_{int,2}$. 
We use eqs. (\ref{overlap}) and (\ref{deltav}), and
we need to calculate the wavefunctions
$| N \rangle , |N + 1 \rangle$ and $| N + 2 \rangle$.
Generically, we can write:
\begin{widetext}
\begin{eqnarray}
\left| \Psi \right\rangle &= &
\left( 1 - \frac{1}{2} \sum_{n_1}
\frac{| \langle n_1 | {\cal H}_{int,1}
| \, 0 \rangle |^2}{( E_0 - E_{n_1} )^2} \right)
\left( \left| 0 \right\rangle  +
\sum_{n_1} \frac{| n_1 \rangle \langle n_1 |
{\cal H}_{int,1} | 0 \rangle}{E_0 - E_{n_1}} +
\sum_{n_2} \frac{| n_2 \rangle \langle n_2 |
{\cal H}_{int,2} | 0 \rangle}{E_0 - E_{n_2}} \right. + \nonumber \\
&+ &\left. \sum_{n_2 , n_1} \frac{| n_2 \rangle \langle n_2 |
{\cal H}_{int,1} | n_1 \rangle \langle n_1 |
{\cal H}_{int,1} | 0 \rangle}{(E_0 -  E_{n_2} )
(E_0 -  E_{n_1} )} + \sum_{n_1}
\frac{| n_1 \rangle \langle n_1 | {\cal H}_{int,1} | 0 \rangle
\langle 0 | {\cal H}_{int,1} | 0 \rangle}
{( E_0 - E_{n_1} )^2} \right)
\label{perturbation}
\end{eqnarray}
\end{widetext}
where $| 0 \rangle , | n_1 \rangle$ and $| n_2 \rangle$
describe states with zero, one and two electron-hole pairs.

The operator $\sum_i  V_i c^\dag_i$ in eq.(\ref{overlap}) 
can, at most, change by one the number of electron-hole
pairs in the wavefunction. Hence, there are no contributions
to eq.(\ref{overlap}) which are of first order in
${\cal H}_{int,2}$. In addition, the terms in the second line
in eq.(\ref{perturbation}) are of second order in ${\cal H}_{int,1}$
and lead to wavefunctions with, at least, one electron-hole
pair. Their contributions are of third order in
${\cal H}_{int,1}$, and need not be considered.
As the basis set used gives the Hartree-Fock solution for
charge state $N$, we have that:
\begin{equation}
{\cal H}_{int,1} | N \rangle_{HF} = 0
\end{equation}
which implies that the corrections to $V_{eff}^{N \rightarrow N+1}$
are of second order in ${\cal H}_{int,1}$.
Finally, adding all contributions, we find:
\begin{widetext}
\begin{equation}
\Delta V \approx \sum_{j<i} \frac{V_j h_{ij}}{\epsilon_j - \epsilon_i}
+ \sum_{k>i} \frac{V_k h_{ik}}{\epsilon_i - \epsilon_k} -
2 V_i \sum_{j \le i}^{k>i} \frac{h_{jk}^2}{( \epsilon_j -
\epsilon_k )^2} + \sum_{j<i}^{k>i} \frac{V_j h_{jk}
h_{ik}}{( \epsilon_i - \epsilon_k ) ( \epsilon_j - \epsilon_k )} 
+ \sum_{j<i}^{k>i} \frac{V_k h_{jk}
h_{ik}}{( \epsilon_j - \epsilon_i ) ( \epsilon_j - \epsilon_k )}
\label{dvperturbation}
\end{equation}
\end{widetext}
where:
\begin{equation}
h_{jk} = \langle j | {\cal H}_{int,1} | k \rangle
\end{equation}
and we assume that all levels are doubly degenerate.

Eq.(\ref{dvperturbation}) implies that electron-electron
interactions within the dot give rise to a first order correction
to the hopping, whose sign depends on the nature of the interaction,
and a second order correction which tends to suppress the hopping.
In the limit where the level spacing within the dot is negligible
with respect to the temperature and charging energy, these
terms change into the non equilibrium corrections to the effective
tunneling density of states analyzed in\cite{UG91,Betal00}. 
Note that in in addition to the correction to the hopping given in 
eq.(\ref{dvperturbation}), the polarization of the occuppied
levels, $j<i$, gives a correction to $V$ which does not depend on the
number of electrons in state $i$. 
\subsection{Spherical dot.}
We now apply the previous analysis to the case of a spherical dot
of radius $R$ and $N \gg 1$ electrons.
For simplicity, we assume that the positive charge needed
to stabilize the system is uniformly distributed within
the radius $R$.
The level spacing is: 
\begin{equation}
\Delta \sim \frac{ \hbar^2 k_F }{ m R }
\label{spacing}
\end{equation}
where
$k_F = ( 9 \pi / 4 )^{1/3} R^{-1}$ is the Fermi wavevector,
and $m$ is the mass of the electrons.
For sufficently large electronic densities,
$k_F \gg ( m e^2 ) / \hbar^2$, the Coulomb repulsion
between electrons can be treated using the
Fermi-Thomas approximation.
Within this approximation,
${\cal H}_{int,1}$ is given by
the Fermi-Thomas potential induced by the addition of a unit of
of charge to the dot. For a three dimensional spherical dot,
this potential is\cite{BMM97}:
\begin{equation}
V_{FT} ( r ) = - \frac{e^2 e^{- k_{FT} ( R - r )}}{k_{FT} ( R - r )^2}
\label{FT}
\end{equation}
where $r$ is the radial coordinate, and $k_{FT} = \sqrt{( 4 e^2 m k_F )
/ ( \pi \hbar^2 )}$ is the Fermi-Thomas wavelength. 
In addition to this term, we have to include a short range
interaction which leads to the excitation of two
electron-hole  pairs when expanding around the 
Hertree-Fock solution, ${\cal H}_{int,2}$ in 
eq.(\ref{dvperturbation}).  
The matrix elements of ${\cal H}_{int,1}$ decay
as $R^{-2}$ for $k_{FT} R \gg 1$, while those
of ${\cal H}_{int,2}$ decay as $R^{-3}$, justifying
the separation of scales used in obtaining
eq.(\ref{dvperturbation}).

The potential $V_{FT}$ is localized within a shell of size
$k_{FT}^{-1}$ around the edges of the dot. We assume that
the matrix elements $\langle k | V_{FT} | j \rangle$ are roughly
constant if $\Delta k \ll k_{FT}$ where
$\epsilon_j - \epsilon_k \approx \hbar v_F \Delta
k$, and zero otherwise. As the level spacing is
given in eq.(\ref{spacing}),
the number of
levels for which the matrix elements of $V_{FT}$ are
not negligible is $k_{FT} R$. Within this energy window,
we can also assume that $| \Psi_k ( {\bf \vec{r}}_0 ) |
\approx | \Psi_i ( {\bf \vec{r}}_0 ) |$, so that the
the correction to $\Delta V$, using
eq.(\ref{dvperturbation}), is the sum of $k_{FT} R$ terms of the
same sign. Finally, $V_{FT}$ conserves angular
momentum.
The maximum angular momentum in a sphere of radius $R$ is
approximately $k_F R$.

Defining:
\begin{equation}
\bar{V}_{FT} = \langle i | V_{FT} | i \rangle
\approx \frac{e^2}{k_{FT}^2 R^3} \sim \frac{\hbar^2}{m k_F R^3}
\label{estimates}
\end{equation}
we find that the first (linear) and second
(quadratic) contributions to
$\Delta V$ are:
\begin{eqnarray}
\Delta V_1 &\propto &V \frac{\bar{V}_{FT}}{\Delta} \log ( k_{FT} R )
\nonumber \\
\Delta V_2 &\propto &V ( k_F R )^2 \frac{\bar{V}_{FT}^2}{\Delta^2}
\log ( k_{FT} R )
\label{dvterms}
\end{eqnarray}
where the $( k_F R )^2$ factor to $\Delta V_2$ arises from the
number of angular momentum channels.
We have obtained that the two corrections
are of similar magnitude, and decay
as $( k_F R )^{-2}$.
\subsection{Diffusive dot.}
The low energy levels and wavefunctions
of quantum dots in the diffusive regime
are well described using random matrix theory. We consider a dot
of  average radius $R$, mean free path $l$, and mean level
separation $\Delta$. In addition, we assume that the electron density
is large, $r_s \ll 1$, where $r_s^{-1} = ( 4 / 9 \pi )^{1/3}
( \hbar^2 k_F ) / ( m e^2 )$. The electron-electron interaction
can be expanded in powers of the inverse conductance, $g = E_T / \Delta$,
where $E_T = ( \hbar^2 k_F l ) / ( m R^2 )$\cite{BMM97,A00,ABG02}.

As in the case considered in the
previous subsection, the leading term in an expansion
in $g^{-1}$ arise from the
Fermi-Thomas potential,
eq.(\ref{FT}), upon the addition of electrons. This potential is
determined solely by the geometry of the dot and electrostatic
constraints, and is independent of the details of the dynamics
of the electrons.

We use the matrix elements of $V_{FT}$ calculated for a
diffusive dot in\cite{BMM97}. As in eq.(\ref{dvterms}),
we find two contributions, $\Delta V_1$ and $\Delta V_2$. 
The average over disorder of $\Delta V_1$ is zero, with
mean fluctuations:
\begin{equation}
\left\langle \Delta V_1^2 \right\rangle_{dis} \propto
V^2 \frac{c_1}{g} \log ( g )
\label{dvterm1}
\end{equation}
where $c_1$ is a dimensionless constant of order unity.
The $\log ( g )$ correction is due to the summation over
states such that $\Delta \le \epsilon_j - \epsilon_k \le E_T$.

To next order in $g^{-1}$, we find:
\begin{equation}
\Delta V_2 \propto V \frac{c_2}{g} \log ( g )
\label{dvterm2}
\end{equation}
For dots of similar radii, $\Delta V$ is larger in the
diffusive regime than in the regular case. For chaotic, ballistic
dots, it seems reasonable to replace $g$ by
$k_F R$ in eqs.(\ref{dvterm1},\ref{dvterm2}).
\section{Mean field analysis.}
\subsection{Magnetic solutions.}
We now consider the hamiltonian in eq.(\ref{hamil}).
We first analyze possible magnetic solutions.
The mean field hamiltonian is:
\begin{widetext}
\begin{eqnarray}
{\cal H}_{MF} &=  &\sum_{k,s} \epsilon_{k} c_{k,s}^\dag c_{k,s}
+ \epsilon^0_d n_d + U \left( \left\langle n_{d \uparrow} \right\rangle
n_{d \downarrow} + \left\langle n_{d \downarrow} \right\rangle
n_{d \uparrow} \right) + V \sum_s \bar{c}^\dag_{0,s} d_s + h. c. \nonumber \\
&- &\Delta V \left(
\left\langle n_{d \uparrow} \right\rangle
\bar{c}^\dag_{0 \downarrow} d_\downarrow +
\left\langle n_{d \downarrow} \right\rangle
\bar{c}^\dag_{0 \uparrow} d_\uparrow +
\left\langle \bar{c}^\dag_{0 \downarrow} d_\downarrow + h. c. \right\rangle
n_{d \uparrow} +
\left\langle \bar{c}^\dag_{0 \uparrow} d_\uparrow + h. c. \right\rangle
n_{d \downarrow} \right)
\label{hamilMF}
\end{eqnarray}
\end{widetext}
where $\bar{c}_0$ is defined in eq.(\ref{dot}).
The sites defined by $c_i , i \ne 0$ can be integrated out. leading to
a $ 2 \times 2$ matrix equation for the components of
the Green's function for
each spin index projected on the $c_0$ and $d$ sites:
\begin{equation}
{\cal G}^{-1}_{ss} ( \omega ) = \left(
\begin{array}{cc} - i N_0^{-1} &- V + \Delta V n_{\bar{s}} \\
- V + \Delta V n_{\bar{s}} &\omega - \epsilon^0_d - 
U n_{\bar{s}} + \Delta V g_{\bar{s}}
\end{array} \right)
\end{equation}
where $s$ is the spin index, and we have defined:
\begin{eqnarray}
n_s &= &\langle n_{d s} \rangle \nonumber \\
g_s &= &\left\langle \bar{c}^\dag_{0 s} d_s + d^\dag_s
c_{0 s} \right\rangle 
\end{eqnarray} 
For the nonmagnetic solution,
the different components of the density of states are (omitting spin
indices):
\begin{eqnarray}
{\rm Im} G_{dd} ( \omega ) &= &\frac{\Gamma}{( \omega - \epsilon_d )^2
+ \Gamma^2} \nonumber \\
{\rm Im} G_{00} ( \omega ) &= &\frac{N_0 ( \omega - \epsilon_d )^2}
{( \omega - \epsilon_d )^2 + \Gamma^2} \nonumber \\
{\rm Im} G_{0d} ( \omega ) &= &- \frac{N_0 ( \omega - \epsilon_d )
( V - \Delta V n_0 )}{( \omega - \epsilon_d )^2 + \Gamma^2}
\label{green}
\end{eqnarray}
where:
\begin{eqnarray}
\Gamma &= &N_0 ( V - \Delta V n_0 )^2 \nonumber \\
\epsilon_d &= &\epsilon^0_d + U n_0 - \Delta V g_0
\end{eqnarray}
The self consistent relations required in eq.(\ref{hamilMF}), and
the Green's functions in eq.(\ref{green}) imply that:
\begin{eqnarray}
n_0 &= &\frac{1}{2} -
\frac{1}{\pi} \arctan \left( \frac{\epsilon_d}{\Gamma}
\right) \nonumber \\
\langle c^\dag_{0 s}
d_s \rangle &= & \frac{N_0 ( V - \Delta V n_0 )}
{2 \pi} \log \left[ N_0^2 (
\epsilon_d^2
+ \Gamma^2  ) \right]
\end{eqnarray}
For the symmetric Anderson model, we have $\epsilon_d
= 0$, and $n_0 = 1/2$.
The equations for the magnetic solution, in the limit of
small magnetization, are analyzed in the Appendix.

\begin{figure}
\resizebox{8cm}{!}{\rotatebox{0}{\includegraphics{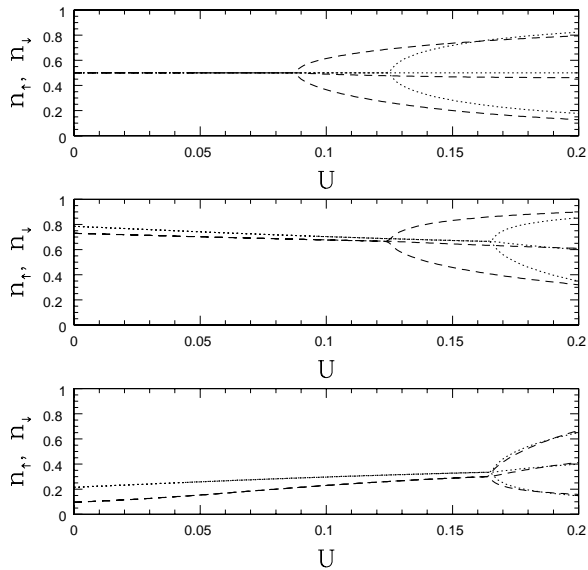}}}
\caption{Number of electrons per spin, as function of $U$. Dotted
curves: $\Delta V = 0$. Broken Curves: $\Delta V = 0.15$.
In all cases, $V  = 0.2$, in units of $N_{\epsilon_F}^{-1}$.
Top: $\epsilon_d^0 + U / 2 
 = 0$. Center:
$\epsilon_d^0 + U / 2 
 = - 0.05$. Bottom:
$\epsilon_d^0 + U / 2  
 = + 0.05$.}
\label{results}
\end{figure}

Numerical results of the occupancies of the different spin states
as function of $U , V , \Delta V$ and $\epsilon_0$ are shown
in Fig.[\ref{results}].
\subsection{Superconducting solutions.}
Alternatively, we can use the BCS decoupling and write a mean field
hamiltonian:
\begin{widetext}
\begin{eqnarray}
{\cal H}_{BCS} &= &\sum_{n,s} t \bar{c}^\dag_{n,s} \bar{c}_{n+1,s} +
\epsilon_d n_d + V \sum_{s} \left( \bar{c}^\dag_{0,s}
d_s + d^\dag_s c_{0,s} \right)
+ U \left\langle d^\dag_\uparrow d^\dag_\downarrow  \right\rangle d_\uparrow
d_\downarrow \nonumber \\
&-  &\Delta V \left[ \left\langle d^\dag_\uparrow
d^\dag_\downarrow  \right\rangle
\left( d_\downarrow \bar{c}_{0 \uparrow} + d_\uparrow c_{0 \downarrow} \right)
+ \left\langle d^\dag_\uparrow \bar{c}^\dag_{0 \downarrow} +
d^\dag_\downarrow \bar{c}^\dag_{0 \uparrow}
\right\rangle d_\uparrow d_\downarrow \right]
+ h. c.
\label{hamilBCS}
\end{eqnarray}
\end{widetext}
This hamiltonian couples up spin electrons with down spin holes, and
viceversa. The Green's function projected on the electron
and hole states at sites $c_0$ and $d$ can be written
in terms of two $4 \times 4$ matrices:
\begin{widetext}
\begin{equation}
{\cal G} ( \omega ) = \left( \begin{array}{cccc}
- i N_0^{-1} &- V &0 &\Delta V n \\
- V &\omega - \epsilon_d &\Delta V n &- U n + \Delta V g\\
0 &\Delta V n &- i N_0^{-1} &V \\
\Delta V n &- U n + \Delta V g&V &\omega + \epsilon_d \end{array}
\right)
\label{green_bcs}
\end{equation}
\end{widetext}
where the two upper (lower) rows correspond to electrons (holes),
and we have defined:
\begin{eqnarray}
n &= &\langle d_\uparrow d_\downarrow \rangle \nonumber \\
g &= &\langle d_\uparrow c_\downarrow +
d_\downarrow c_\uparrow \rangle
\label{parameters_BCS}
\end{eqnarray}
The selfconsistency condition implicit in eq.(\ref{green_bcs}) can be
linearized, as outlined in the Appendix.
\begin{figure}
\resizebox{7cm}{!}{\rotatebox{0}{\includegraphics{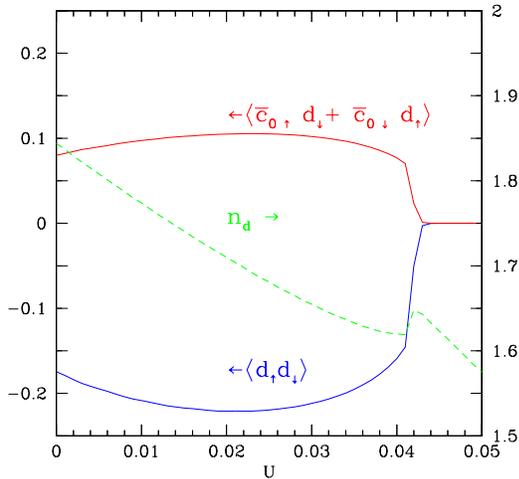}}}
\caption{BCS order parameters, eq.(\protect{\ref{parameters_BCS}}),
 for $V = 0.2 , \Delta V
= 0.15$, and $\epsilon_d - U/2 = 0$ in units of $N ( \epsilon_F )^{-1}$.
The transition is weakly first order. Broken line: occupancy of
the localized level (right scale).}
\label{bcs}
\end{figure}
\begin{figure}
\resizebox{7cm}{!}{\rotatebox{0}{\includegraphics{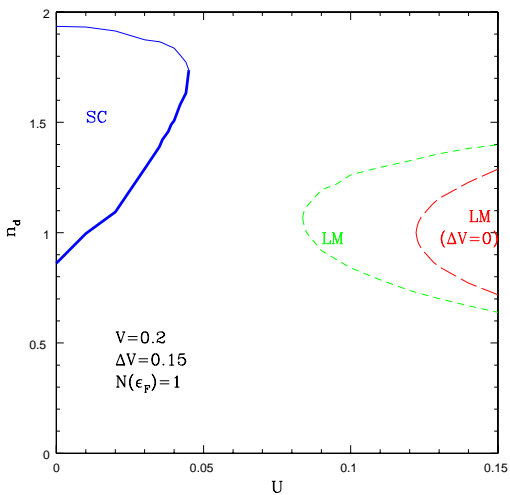}}}
\caption{Mean field phase diagram of the hamiltonian
 in eq.(\protect{\ref{hamil}}). SC: superconducting solution.
The thick line is a first order transition
(see Fig.[\protect{\ref{bcs}}]). The jump in
$n_d$ is comparable to the thickness of the line.
LM: local moment. For comparison, the results for
$\Delta V = 0$ are also plotted.}
\label{phase_d}
\end{figure}
\subsection{Corrections beyond the mean field approximation.}
The impurity model studied here can be considered as a (0+1) dimensional
model, where fluctuations in time need to be considered. In principle,
we can integrate out the fermion in the leads, and obtain an effective
four state model. The four states correspond to the four
occupancies of the single state at the impurity, as defined
in eq.(\ref{hamil}). These states have 
long range interactions in time, as  in
related fermionic and bosonic problems\cite{AY71,GHM85}.
The fluctuations in time will tend to restore
the symmetries broken at the mean field level. We will use the
mean field results as a guide to infer which type of
universality class will determine the low temperature properties.
The fact that the reduced Hilbert space associated to the impurity
includes only four states 
greatly restricts the possible
low energy fixed points.

The analysis in the two previous subsections suggests that
there are two physical regimes: i) Near $n_d = 1$ and
for large values of $U / \Gamma$, a local magnetic moment is induced, and, ii)
away from $n_d = 1$, and for moderate values of $U / \Gamma$,
pairing correlations develop near the dot or impurity.

We identify the local moment regime studied earlier with the Kondo effect.
Note, however, that
the relation between the Kondo temperature and the
parameters in the hamiltonian will be changed by the 
assisted hopping term. 

Pairing fluctuations imply that charge states
which differ by two are strongly mixed in the ground state. This can be
checked by the exact diagonalization of a block containing
the impurity and a few 
sites in the leads. The assisted hopping term
leads to an effective negative $U_{eff}$ model, where 
$U_{eff} = E_N + E_{N+2} - 2 E_{N+1}$, for certain values of $N$.
These relatively small blocks, when attached to the rest of the 
leads will act as local defects with a tendency towards
pairing.
Hence, we conclude that the BCS solution identified in
the mean field approximation corresponds to a regime 
described by the negative $U$ Anderson model.
\section{Conclusions.}
We have analyzed the influence of assisted hopping in 
correlated impurities embedded in a metal, or quantum dots
coupled to leads. This type of interaction is the simplest
term which can be defined in the restricted basis given by 
a single quantum level within the dot. Related effects, in the
opposite limit where the states within the dot can be treated
as a continuum were studied in\cite{UG91,Betal00}.

We have estimated, in section II, the magnitude of this term for
simple models of regular and diffusive dots. In the latter case,
we find that the assisted hopping term has a similar dependence
on the conductance of the dot as the interaction corrections 
to the peak spacing\cite{BMM97,AG01,UB01,UB02,AM02}. The assisted hopping term
is directly related to the changes of the wavefunctions 
within the dot upon the addition of a single electron.
This interaction will be larger than estimated in section II
in one and two dimensional geometries as the Fermi-Thomas potential is more
extended, and for low electronic
density\cite{WMG99}. Note also that the orthogonality catastrophe is expected
to be enhanced in disordered two dimensional systems\cite{Getal02}.
We have only taken into account the interactions within the dot. 
In devices with more than one dot, the interactions between electrons 
in different dots will also enhance the effects reproted here. 
Assisted hopping can contribute to change the peak height 
distribution with respect to that predicted by Random Matrix
Theory\cite{JSA92,AW01}. 

Our results suggest that assisted hopping can lead to local pairing
correlations for reasonable values of the parameters. This implies 
enhanced conductivity through the dot or impurity, significant
deviations from the electron-hole symmetry implicit in the Kondo
effect, and measurable differences in the conductance through the
two peaks associated to the same quantum level. The
existence of these effects does not seem incompatible with
present experimental evidence\cite{Setal00}. A detailed analysis 
of assisted hopping effects
quantum dots, in the limit of large level spacing,
deserves further investigation.
\section{Acknowledgements.}
Many productive suggestions from J. Hirsch
are gratefully acknowledged.
I am thankful to K. Held, for a careful 
reading of the manuscript.
Financial support from CICyT (Spain), through
grant no. PB96-0875 is gratefully acknowledged.
\section{Appendix.}
\subsection{Linearization of the magnetic Hartree-Fock equations.}
We first analyze the magnetic solutions when
the magnetization is small. Then, we can write:
\begin{eqnarray}
n_{d \uparrow \downarrow} &\approx &n_0 \pm \delta n \nonumber \\
g_{\uparrow \downarrow} &\approx &g_0 \pm \delta g
\label{occup}
\end{eqnarray}
The Hartree Fock approximation introduces four variational
parameters, $\epsilon_{d \uparrow} , \epsilon_{d \downarrow} ,
\Gamma_{\uparrow}$ and $\Gamma_{\downarrow}$. We also make the
expansion:
\begin{eqnarray}
\epsilon_{d \uparrow \downarrow} &\approx &\epsilon_0 \pm \delta
\epsilon \nonumber \\
\Gamma_{\uparrow \downarrow} &\approx &\Gamma_0 \pm \delta
\Gamma
\label{param}
\end{eqnarray}
The consistency between eqs.(\ref{occup}) and eqs.(\ref{param}) leads
to:
\begin{widetext}
\begin{eqnarray}
\delta n &\approx &\frac{\Gamma_0 \delta \epsilon}{\pi ( \Gamma_0^2
+ \epsilon_0^2 )} - \frac{\epsilon_0 \delta \Gamma}
{\pi ( \Gamma_0^2 + \epsilon_0^2 )} \nonumber \\
\delta g &\approx &\frac{N_0 \Delta V \delta n}{2 \pi}
\log \left( \frac{W^2}{\Gamma_0^2 + \epsilon_0^2} \right) +
\frac{N_0 ( V - \Delta V n_0 ) ( \epsilon_0 \delta \epsilon
+ \Gamma_0 \delta \Gamma )}{\pi ( \Gamma_0^2 + \epsilon_0^2 )} \nonumber \\
\delta \epsilon &\approx &- U \delta n + \Delta V \delta g \nonumber \\
\delta \Gamma &\approx &2 N_0 V \Delta V \delta n
\label{linear}
\end{eqnarray}
Eqs. (\ref{linear}) have a non trivial solution if:
\begin{equation}
\rm{Det} \left| \begin{array}{cccc}
1 &U &0 &-\Delta V \\ - \frac{\Gamma_0}{\pi ( \Gamma_0^2 + \epsilon_0^2 )}
&1  &\frac{\epsilon_0}{\pi ( \Gamma_0^2 + \epsilon_0^2 )} &0\\
0 &-2 N_0 V \Delta V &1 &0 \\
- \frac{N_0 ( V - n_0 \Delta V ) \epsilon_0}
{\pi ( \Gamma_0^2 + \epsilon_0^2 )} &- \frac{N_0 \Delta V}{2 \pi} \log \left(
\frac{W^2}{\Gamma_0^2 + \epsilon_0^2} \right) &
- \frac{N_0 ( V - n_0 \Delta V ) \Gamma_0}{\pi ( \Gamma_0^2 + \epsilon_0^2 )}
&1 \end{array} \right|
\label{det}
\end{equation}
\end{widetext}
For $\Delta V = 0$, eq.(\ref{det}) reduces to:
\begin{equation}
\frac{U \Gamma_0}{\pi ( \Gamma_0^2 + \epsilon_0^2 )}
= 1
\end{equation}
For $\epsilon_0 = 0$ (symmetric case), and $\Delta V \ne 0$, we
find:
\begin{equation}
U = \pi \Gamma_0 - \frac{N_0 \Delta V^2}{\pi}
\left[ \frac{2 V}{V - n_0 \Delta V} + \log \left( \frac{W}{\Gamma_0}
\right) \right]
\end{equation}
This equation defines the critical value of $U$ at which a solution with
a non zero magnetization appears.

\subsection{Linearization of the BCS equations.}
We define
$\tilde{G}_{dd} ( \omega ) $ and $\tilde{G}_{d c_0} ( \omega )$
as the anomalous Green's functions involving electrons and holes
induced by the BCS coupling in eq.(\ref{hamilBCS}). They can be obtained
from the implicit equation (\ref{green_bcs}). The selfconsistency
equations are:
\begin{eqnarray}
n &= &\frac{1}{\pi} \int_{- \infty}^0 {\rm Im} \tilde{G}_{dd} ( \omega )
d \omega \nonumber \\
g &= &\frac{1}{\pi} \int_{- \infty}^0 {\rm Im} \tilde{G}_{d c_0}
 ( \omega ) d \omega
\end{eqnarray}
We can invert eq.(\ref{green_bcs}) and linearize with respect to
$n$ and $g$, to obtain:\begin{widetext}
\begin{eqnarray}
{\rm Im} \tilde{G}_{dd} ( \omega ) &= &
\frac{2 V \Delta V N_0 ( \omega^2 - \epsilon_d^2 - \Gamma^2 ) n
+ 2 \Gamma \omega ( U n - \Delta V g )}
{\left[ ( \omega - \epsilon_d )^2 + \Gamma^2 \right]
\left[ ( \omega + \epsilon_d )^2 + \Gamma^2 \right]} \nonumber \\
{\rm Im} \tilde{G}_{d c_0} ( \omega ) &= &
- \frac{2  \Gamma^2 \Delta V N_0 \omega n +
( \omega^2 - \epsilon_d^2 - \Gamma^2 )
\left[ N_0 \Delta V ( \omega - \epsilon_d ) n - N_0 V ( U n - \Delta V g )
\right] }
{\left[ ( \omega - \epsilon_d )^2 + \Gamma^2 \right]
\left[ ( \omega + \epsilon_d )^2 + \Gamma^2 \right]}
\end{eqnarray}
We can integrate these expressions, so that:
\begin{eqnarray}
n &= &- \frac{U n - \Delta V g}{\pi \epsilon_d} \arctan
\left( \frac{\epsilon_d}{\Gamma} \right) \nonumber \\
g &= &- \frac{ 2 N_0 \Delta V n}{\pi}
\left\{ \log \left[ N_0^2
( \epsilon_d^2 + \Gamma^2 ) \right] + \arctan
\left( \frac{\epsilon_d}{\Gamma} \right) \right\}
\label{param_bcs}
\end{eqnarray}
\end{widetext}
Finally, we obtain:
\begin{widetext}
\begin{equation}
\left[ 1 + \frac{U}{\pi \epsilon_d} \arctan
\left( \frac{\epsilon_d}{\Gamma} \right) \right]
+ \frac{2 N_0 \Delta V^2}{\pi^2 \epsilon_d}
\arctan
\left( \frac{\epsilon_d}{\Gamma} \right)
\log \left[ N_0^2 ( \epsilon_d^2 + \Gamma^2 ) \right] \le 0
\label{critical}
\end{equation}
\end{widetext}
where we have dropped the $\arctan$ term in
the second equation in (\ref{param_bcs}).

\end{document}